\documentstyle[preprint,aps]{revtex}
\begin{document}
\draft
\title{QHE, magnetoresistance and disordered transport on 2D mesoscopic
plaquettes.}
\author{A.~Aldea, P.~Gartner,and M.~Ni\c{t}\~a}
\address{Institute of Physics and Technology of Materials, POBox MG7,
Bucharest-Magurele, Romania}
\maketitle
\begin{abstract}
The transport properties of a rectangular mesoscopic
plaquette in the presence of a perpendicular magnetic field are studied
in a tight-binding model with randomly distributed traps. The longitudinal
and Hall resistances are calculted in the four-probe Landauer-B\"{u}ttiker
formalism which accounts automatically both for the quantum coherence and
the trapping-induced localization. The localized character of eigenvectors
and the specific aspect of the density of states at a given magnetic
flux are correlated with the behaviour of the mentioned resistances as
function of the Fermi energy. The Hall insulator and quantum Hall regimes 
are evidentiated.  The dependence on magnetic field of the configurational 
averages of the 
longitudinal and Halll resistance is studied in a purely
quantum-mechanical approach. Both negative and positive magnetoresistances
are found.
\end{abstract}


The electronic mesoscopic transport on a plaquette in the presence of an
external random atractive potential (traps) and of a perpendicular magnetic 
field exhibits a number of interesting properties. They are due to the 
modifications induced by traps in the Hofstadter energy spectrum and to 
the changes in the nature of the eigenstates. For a clean, infinite 
plaquette the spectrum consists
of alternating bands and magnetic gaps described by the Hofstadter butterfly.
In a finite plaquette edge states start filling the gaps (thus turning them 
into quasi-gaps). The presence of traps, beside shifting the spectrum 
downwards, gives rise to localized states at its bottom. When the traps are 
sufficiently deep and rare, the localized levels are well separated by a 
spectral gap from the rest of the spectrum which is still reminiscent of 
the former butterfly. 
This case turns out to be the most interesting one since band-, localized-, 
and edge-states  are encountered
simultaneously with true and pseudo-gaps.

The transport properties of the plaquette should show at least two distinct 
behaviours: a high resistive one, for energies corresponding to either 
localized
states or the gap, and the quantum Hall behaviour, when the transport takes
place along the edge states in the pseudo-gap. It is hard to say, a priori, how
the Hall resistance will be looking like in the first case and also what the
effect of traps on the edge-state transport will be. These aspects will be
discussed below.

In the present frame, new arguments can be obtained concerning the problem of
interference effect on the conduction phenomena in weak magnetic field and 
strongly localized regime. Unlike other approaches [1-3] we are able to take 
completely into consideration the quantum coherence (the backscattering is 
included)
and to describe the Hall resistance,too. In the mentioned papers, the magnetic 
field
dependence of the transition probabilities between two quantum states appears 
exclusively through the resonance integral. In finite samples, the resonance 
integral shows strong mesoscopic oscillations as the magnetic field increases 
from zero, however a configurational average will have a smooth dependence. 
Performing 
a 'logarithmic average' over all directed paths (contained in a cigar-
shaped region between two states),~ Ngyen,Shklovskii and Spivak [1] get a
negative magnetoresistance, linear in magnetic field. The same result is
obtained by Schirmacher in an effective medium approximation [2], but, using
again EMA, B\"{o}ttger et al.[3] get a quadratic behaviour. Using another
way of averaging, Entin-Wohlman at al.[4] obtain a quadratic dependence in 
the low field limit, both negative or positive, depending on the 
resistivity of the
system. The experiments show  two types of behaviour,too [5].

The longitudinal and Hall resistances of a plaquette with four attached leads 
are calculated at zero temperature using the Landauer-B\"{u}ttiker  formalism
and a tight-binding Hamiltonian. The conductances $g_{\alpha\beta}$ between
two leads denoted by $\alpha$ and $\beta$ are given by the scattering 
amplitude between the leads
and the resistances $R_{L}$ (longitudinal) and $R_{H}$ (Hall) can be expressed
in terms of $g_{\alpha\beta}$ [6]. The details can be found in [7] and
a similar approach was used by Gagel and Maschke in Ref.8. Since in
2D the strong localization cannot be induced by disorder, we shall induce it
by trapping. In the tight-binding model, the traps are simulated by negative 
energies assigned to some sites in the diagonal term of the Hamiltonian 
describing the plaquette :
\begin{eqnarray}
&&H^{S}=\sum_{n=1}^{N}\sum_{m=1}^{M} [~\epsilon_{nm} \vert n,m
\rangle\langle n,m\vert
+ e^{i2\pi\phi m}\vert n,m\rangle\langle n+1,m\vert
+\vert n,m\rangle\langle n,m+1\vert + h.c.~] \,,
\end{eqnarray}
where ${\vert n,m \rangle}$ is a set of orthonormal  states localized at the
sites (n,m), and $\phi$ is  the magnetic flux through the unit cell
(measured in quantum flux units.). In Eq.(1) $\epsilon_{nm}=0$ except for
the trap sites which are chosen at random with concentration $c$. The
hopping term is taken equal to 1 (up to the magnetic phase) and this specifies
the energy scale.

The integrated density of states (IDOS) of this Hamiltonian is shown in Fig.1
The black regions represent the regions where the eigenvectors are grouped
together forming bands. In the grey regions, IDOS increases slowly with the
energy, the eigenvalues are rare and the corresponding eigenvectors are 
localized on the edges. It can be checked that below the lowest band there are
few eigenvalues , the corresponding vectors being strongly localized on the 
traps. Further information can be obtained from the degree of occupation
of the traps, i.e.
$\sum_{i} |{\Psi}_i|^{2}, (i\in traps)$. This quantity  is shown for different 
energies in Fig.2. It takes high values for localized states as it should, 
however small
maxima appear also in pseudo-gaps due to the admixture of  
impurity states localized in the pseudo-gaps with the 
quasi-continuum of the edge states. 

At the bottom of the spectrum, as far as the energy runs through the domain
of localized states, the Hall resistance shows oscillations. However, once
entering the gap a Hall insulator behaviour is installed. This is characterized
by extremely high values of  $R_{L}$, but also by small values and a smooth 
dependence on energy of $R_{H}$ . The Hall insulator regime is  explained 
easily along with the argument outlined in [9]. Essentially it arises from 
the fact 
that the propagator between lead contact points is exponentially small
$g_{\alpha,\beta} \sim {\lambda}^2$ where
 $\lambda\propto~ exp{(-\xi d_{\alpha\beta})}$~
($\xi=$ the inverse decay length, $d_{\alpha\beta}$ =distance between 
contacts). The leading ${\lambda}^2$-term of the conductance matrix is symmetric 
and the antisymmetric part starts with the next-to-leading ${\lambda}^4$-term.
One is lead to the conclusion that the 
inverse of $g_{\alpha\beta}$ ( which gives the resistances) contains 
a singular ${\lambda}^{-2}$ part, which is symmetric.
Therefore $R_{H}$ ,being an antisymmetric quantity, remains non-singular.

The first QH plateau appears above the Hall insulator domain. In the QH regime
there is a competition between the localization at the edges due to the magnetic 
field and the 
localization on the trapping sites due to the electrostatic 
atractive potential as discussed above.  
In spite of this one has to note that the
plateaus in Fig.3 are not affected by this change in the nature of the 
eigenfunction. Of course, at high trap concentration, the localization on 
traps is much stronger, the pseudo-gaps and the plateaus disappear
at the inferior part of the spectrum, which is more influenced by the
trapping potential, however they may be maintained at higher
energies.

The average values of $R_{L}$ and $R_{H}$ over an ensemble of mesoscopic
plaquettes may give hints on the expected macroscopic behaviour of 
these entities. The macroscopic condition of self-averaging  is difficult 
to be reached technically because it
requires a large plaquette. For the dimension we used (64x10 atomic distances) 
the distribution function of $R_{L}$ is broad and consequently
$\langle R_{L}\rangle\neq\langle 1/ R_{L}\rangle^{-1}$. Nevertheless 
the averages
$\langle R_{L}\rangle, \langle log R_{L}\rangle$ and $\langle R_{H}\rangle$
show a smooth dependene on the magnetic flux  and this is true
for the second moment, too. These average values have been calculated 
at two different energies, corresponding to the localized and band regime.
In the localized case, $\langle R_{L}\rangle$ and $\langle log R_{L}\rangle$
show negative magnetoresistance with a $B^{2}$-dependence at low magnetic 
field $B$ as in [3,4]. This dependence is weakened gradually with increasing 
$B$, ending up with the change of slope at higher fields (see Fig.4). 
On the contrary,
in the band regime (see Fig.5), we found a linear positive magnetoresistance 
at low fields, followed by a change to negative behaviour at higher fields.
In the present approach, one realizes why in the strongly localized regime
(hopping) the magnetoresistance should be quadratic. This stems from the
dominance (see above) of the symmetric part of $g_{\alpha \beta}$ ,which is an
even function of $\phi$. No such arguments hold in the band regime. 
In what concerns the Hall resistance,
it starts linearly, but changes eventually the slope and even the sign. 

In conclusion, the mesoscopic plaquette with random traps allows the study 
of the  transport properties both in the localized and quantum Hall regime.
In the range of localized states, one observes the Hall insulator 
behaviour. At higher 
energies, QH plateaus exist even in the presence of traps, in spite of the
fact that the edge states get some localization on the trapping sites.
For ensemble-averaged values, this purely quantum-mechanical approach confirms 
the negative 
magnetoresistance (starting with a quadratic dependence on 
the magnetic field) in the case of strong localization. The Hall resistance
could also be calculated, but a deeper scrutiny of both localized and
band case is necessary.

Acknowledgement. The authors are very much indebted to Professor J\'{a}nos 
Hajdu for suggesting this topic.

\begin{figure}
\caption{The integrated density of states of a disordered plaquette ($\phi =0.1,
~ E_{trap}=-3.0,~ c=1/64$).}
\label{fig1}
\end{figure}
\begin{figure}
\caption{The degree of localization on traps  $\sum_{i}|{\Psi}_i|^2(i\in traps)$
as function of energy for the  same plaquette as in Fig.1.}
\label{fig2}
\end{figure}
\begin{figure}
\caption{The Hall resistance as function of energy for the same plaquette as in
Fig.1.}
\label{fig3}
\end{figure}
\begin{figure}
\caption{The ensembe averages $\langle R_{l}\rangle$ and $\langle R_{H}\rangle$ 
as function of the magnetic flux in the high resistance regime ($E_{Fermi}=-4.2,~ 
E_{trap}=-3.0,~ c=1/64$)}
\label{fig4}
\end{figure}
\begin{figure}
\caption{The same as in Fig.4 in the low resistance regime ($E_{Fermi}=-0.5$)}
\label{fig5}
\end{figure}

\begin{references}
\bibitem{1}
V.L.NGUEN,B.Z.SPIVAK,B.I.SHKLOVSKII, Sov.Phys.JETP {\bf 62},1021 (1985).
\bibitem{2}
W.SCHIRMACHER, Phys.Rev.{\bf41},2461 (1990)
\bibitem{3}
H.B\"{O}TTGER,V.V.BRYKSIN and F.SCHULZ,Phys.Rev.B {\bf 49},2447 (1994).
\bibitem{4}
O.ENTIN-WOHLMAN,Y.IMRY,U.SIVAN, Phys.Rev.{\bf B40},8342,(1989).
\bibitem{5}
C.H.OLK,S.M.YALISOVE,J.P.HEREMANS and G.L.DOLL, Phys.Rev.B 
{\bf 52},4643 (1995) ;
L.ESSALEH,J.GALIBERT,S.M.WASIM and E.HERNANDEZ, Phys.Rev. B 
{\bf52},7798 (1995).
\bibitem{6}
M. B\"{U}TTIKER, Phys. Rev. B {\bf 38},9375 (1988).
\bibitem{7}
A.ALDEA,P.GARTNER,A.MANOLESCU and M.NITA, Phys.Rev.B {\bf55},R13389 (1997). 
\bibitem{8}
F.GAGEL and K.MASCHKE, Phys.Rev.B {\bf52},2013 (1995).
\bibitem{9}
P. GARTNER and A. ALDEA, Z. Physik B {\bf 99},367 (1996).
\end{references}
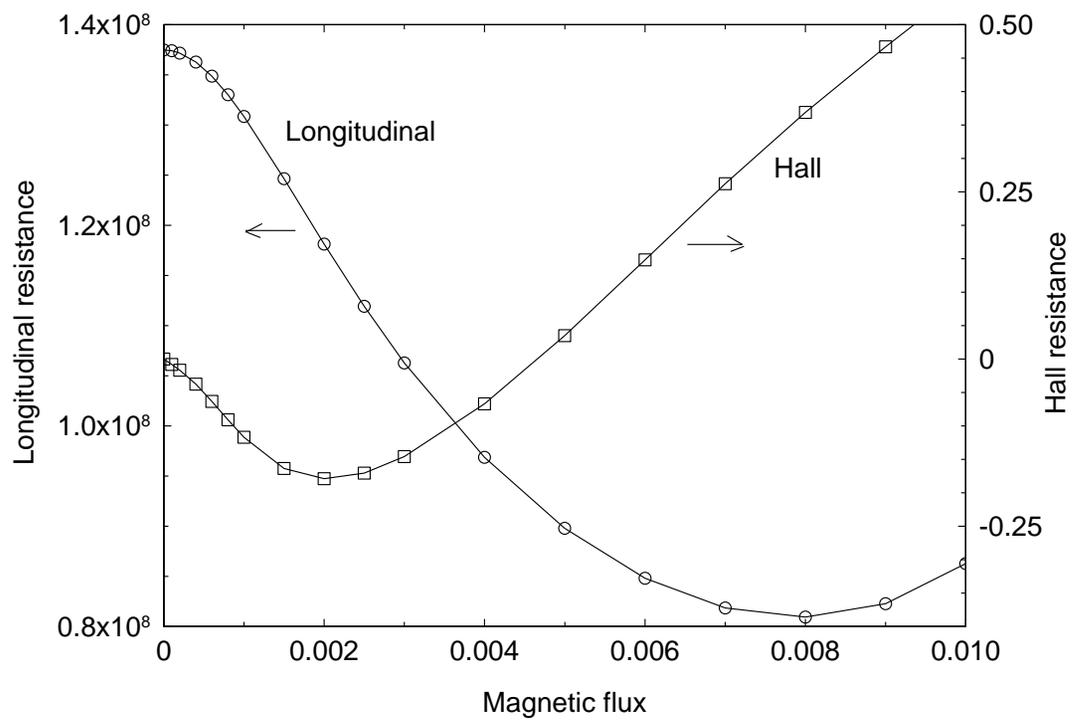
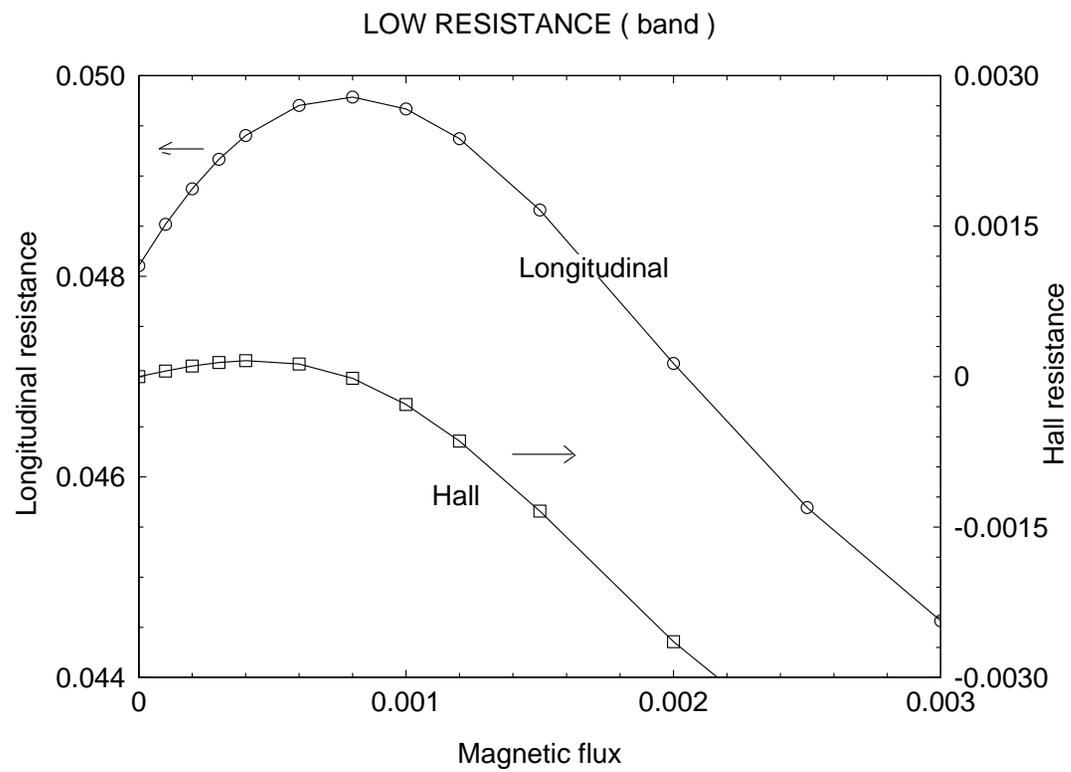
\end{document}